\documentclass{article}
\usepackage{spconf,amsmath,epsfig}
\usepackage[ansinew]{inputenc}
\usepackage{amssymb}
\usepackage{amscd}
\usepackage{eucal}
\usepackage{amsmath}
\usepackage{graphicx}
\usepackage{subfigure}

\title{Cooperative spectrum sensing over unreliable reporting channel}
\name{Amanda de Paula, Cristiano Panazio}
\address{University of São Paulo \\ Escola Politécnica
\\
Email:  \{amanda,cpanazio\}@lcs.poli.usp.br}
\begin{document}
%
\maketitle
\begin{abstract}
This article aims to analyze a cooperative spectrum sensing scheme using a centralized approach with unreliable reporting channel. The spectrum sensing is applied to a cognitive radio system, where each cognitive radio performs a simple energy detection and send the decision to a fusion center through a reporting channel. When the decisions are available at the fusion center, a $n$-out-of-$K$ rule is applied. The impact of the choice of the parameter $n$ in the cognitive radio system performance is analyzed in the case where the reporting channel introduces errors.
\end{abstract}
\begin{keywords}
cognitive radio, cooperation, spectrum sensing, data fusion.
\end{keywords}
\section{Introduction}
\label{sec:intro}

The increasing demand for communication resources is leading to a scarcity in the spectral bands available to transmission. Such scarcity is mainly due to the inflexible spectrum utilization regulamentation, where the bands are statically allocated. As shown in \cite{Ellingson05_02}, this statical spectrum allocation leads to an inefficient spectral occupancy.

Motivated by the necessity of implementing more efficient band allocation schemes, several papers have recently proposed systems based on cognitive radio (CR) \cite{Mitola99}, \cite{Haykin05}. In such systems, secondary users (SU) are allowed to occupy the band licensed to primary users (PU), if the PU are not using the spectral band for that time.

Therefore, the SU must be able to determine whether the spectral band is free or not. This task is accomplished by performing spectral sensing, which can be implemented with several types of algorithms \cite{Mishra09}, \cite{Haykin09}, \cite{Li09}, where the simplest approach is by the means of an energy detection. The main advantage of this spectrum sensing scheme is that it does not require a high \emph{a priori} knowledge about the PU signal. On the other hand, it does not provide a good performance when compared to other techniques, such as feature and coherent detection \cite{sayed08}. An alternative to improve the energy detector performance is applying cooperative algorithms \cite{sayed08}, \cite{letaief09_2}, \cite{ghasemi07}. These cooperative algorithms bring the possibility to combine the measurements provided by the various cognitive radios in the system in order to generate a more reliable spectral sensing.

The cooperative spectrum sensing can be performed by the exchange of soft information \cite{geoffrey_li08} or quantized hard information \cite{ghasemi07}. It is often interesting to implement cooperative cognitive radio system applying hard decision in order to simplify the exchange of information between the cognitive radios and the fusion center. Restricting our attention to this case, a problem that arises is how to merge the decisions provided by the different cognitive radios in order to provide a more reliable sensing.

In \cite{ghasemi07} and \cite{Letaief09} is pointed out that the OR decision rule is more suitable in many cases of practical interest. However, these analysis considered that the reporting channel between the cognitive radio and the fusion center was perfect. Restricting the decision rule to the OR rule, \cite{Letaief08} investigated the effect of reporting errors introduced in the system.

In this article, we will assume the same context in \cite{Letaief08}, but we will investigate the decision rules of the kind $n$-out-of-$K$, observing that, differently from the perfect reporting channel situation, the decision rule which provides the best system performance is not the OR, \emph{i.e.} the 1-out-of-$K$ rule.

This article is organized as follows. In Section \ref{sec:system_model}, the system model utilized throughout this paper is depicted. In Section \ref{Sec:problem_form}, local and cooperative spectrum sensing are described. Section \ref{Sec:results} presents theoretical and simulated results. Finally, in Section \ref{Sec:conclusions}, the conclusions of the paper are stated.

\section{System Model}
\label{sec:system_model}
\begin{figure}[h]
\centering
\includegraphics[width=70mm,height=55mm]{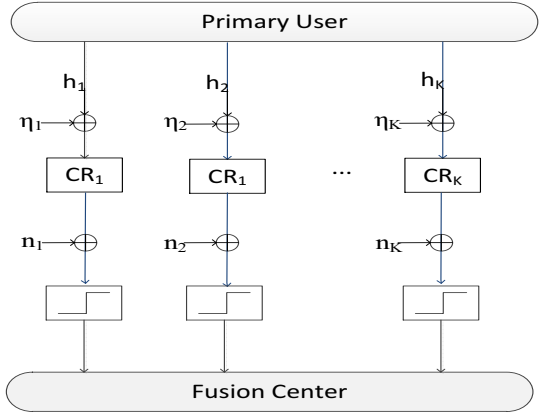}
\caption{System Model}
\label{System_model}
\end{figure}

In this article, we consider a cooperative cognitive radio system with $K$ SU. As depicted in Fig. \ref{System_model}, we assume that the $i^{th}$ cognitive radio receives the signal transmitted by the PU through a channel $h_i$ and that the signal is corrupted by additive white Gaussian noise (AWGN). Each cognitive radio senses the spectrum using an energy detector and sends its one-bit quantized decision to the fusion center.  The signal received by the fusion center sent by each cognitive radio is corrupted with AWGN noise with variance $\sigma^2_{n_i}$.

Finally, the spectral sensing is performed in the fusion center, where a $n$-out-of-$K$ rule is applied, \emph{i.e.}, the fusion center states that the PU is active if the received decision is sent by at least $n$ out of the $K$ cognitive radios.

\section{Problem Formulation}
\label{Sec:problem_form}
\subsection{Local Sensing}
The received signal in the $i^{th}$  cognitive radio can be expressed as one of the following hypothesis:
\begin{equation}
r(n)=\begin{cases}
h_ix(n)+\eta_i(n), \; \quad &\mathcal{H}_0 \\
\eta_i(n), \; \qquad &\mathcal{H}_1
\end{cases}, \quad 1\leq n \leq M
\end{equation}
where $h_i$ is the channel coefficient, which is assumed to be a complex Gaussian random variable, $x(n)$ is the signal transmitted by the PU and $\eta_i(n)$ is AWGN signal with variance $\sigma_{\eta_i}^2$.

Each cognitive radio will apply an energy detection rule in order to decide between these two hypothesis. This decision rule consists in the comparison of the estimated signal energy to a given threshold $\lambda$. The estimated received signal energy, \emph{i.e.} the decision statistic, is given by:

\begin{equation}
T(r)=\frac{1}{\sigma_{\eta_i}^2}\sum_{n=1}^{M} \left|r(n)\right|^2
\end{equation}

The hypothesis test is them accomplished by:
\begin{equation}
T(r)\gtrless^{\mathcal{H}_1}_{\mathcal{H}_0}\lambda
\end{equation}

This means that the $i^{th}$ cognitive radio will state that the spectrum is occupied by the PU if the metric $T(r)$ is greater than $\lambda$.

In the specification of spectrum sensing systems, two parameters are extremely relevant. One of them is the false alarm probability ($P_f$), which is defined as the probability of the cognitive radio declares that the spectrum is occupied under $\mathcal{H}_0$, \emph{i.e.}:
\begin{equation}
P_f=\textrm{Pr}\left\{T(r)\geq\lambda | \mathcal{H}_0\right\}
\end{equation}

This probability measures the efficiency of the cognitive radio system radio, given that if the system presents a low $P_f$ it means that the spectrum holes are allowed to be occupied by the CR more often.

The second important parameter in the cognitive radio system is the miss detection probability ($P_m$), that is defined as the probability of the cognitive radio states that the spectrum is free given that the PU is transmitting:
\begin{equation}
P_m=\rm{Pr}\left\{T(r)<\lambda | \mathcal{H}_1\right\}
\end{equation}

For a single cognitive radio in a fading scenario, these probabilities have been derived in \cite{digham07} and can be expressed as:

\begin{equation}
P_f=\frac{\Gamma\left(M,\frac{\lambda}{2}\right)}{\Gamma\left(M\right)}
\end{equation}

\begin{eqnarray}
P_m=e^{-\frac{\lambda}{2}}\sum_{l=0}^{M-2}{\frac{\left(\frac{\lambda}{2}\right)^l}{l!}}+\left(\frac{1+\gamma}{\gamma}\right)^{M-1} \nonumber\\
\times \left(e^{-\frac{\lambda}{2+2\gamma}}-e^{-\frac{\lambda}{2}}\sum_{l=0}^{M-2}{\frac{\left(\frac{\lambda \gamma}{2+2\gamma}\right)^l}{l!}}\right)
\end{eqnarray}
where $\Gamma(x)$ is the gamma function, $\Gamma(x,y)$ is the upper incomplete gamma function and $\gamma$ is the average system signal-to-noise ratio (SNR) per sample under $\mathcal{H}_1$.

It is important to note that the $P_f$ and $P_m$ are parameterized by the threshold $\lambda$. $P_f$ is a decreasing function of $\lambda$, while $P_m$ is an increasing function of $\lambda$. Therefore, in order to specify the threshold $\lambda$, one should analyze the compromise between low $P_f$ and high $P_m$.

In \cite{letaief09_2} the system parameters were optimized in order to minimize the total error, \emph{i.e.}, $P_f+P_m$. Another common approach to determine the system parameters is the following: for a given $P_m$, determine what are the system parameters that lead to the lower $P_f$ \cite{sayed08}. This approach provides the highest spectrum occupancy given the PU is protected under a specified $P_m$.

\subsection{Cooperative sensing}
Previously, we have analyzed the spectrum sensing performed in each cognitive radio. In this subsection, we deal with the data processing in the fusion center.

We will consider that the $i^{th}$ cognitive radio sends a one-bit decision to the fusion center and that the channel between the cognitive radio and the fusion center is corrupted by an AWGN signal:
\begin{equation}
s_i=d_i+n_i
\end{equation}
where $d_i=\left\{0,1 \right\}$ is the decision sent by the $i^{th}$ cognitive radio and $n_i \sim \mathcal{N}\left(0,\sigma_{n_i}^2\right)$.

Furthermore, the error in the $i^{th}$ cognitive radio is given by:
\begin{equation}
P_e^i=Q\left(\frac{1}{2}\sqrt{\frac{1}{\sigma_i^2}}\right)
\end{equation}
where $Q(x)$ is the complementary error function.

In this article, in order to simplify the analysis, we will consider that the cognitive radio's reporting channel present the same SNR ($\sigma_i=\sigma, \; i=1\dots K$).

Applying the $n$-out-of-$K$ rule, we have that the false alarm and miss-detection probabilities after the decision provided by the fusion center are given by:
\begin{eqnarray}
\label{eq:Qf}
Q_f=\sum_{i=0}^{K-n}{{K}\choose{i}}\left[\left(1-P_f\right)\left(1-P_e\right)+P_fP_e\right]^i \nonumber\\
\times \left[P_f\left(1-P_e\right)+\left(1-P_f\right)P_e\right]^{K-i}
\end{eqnarray}
\begin{eqnarray}
\label{eq:Qm}
Q_m=\sum_{i=K-n+1}^{K}{{K}\choose{i}}\left[P_m\left(1-P_e\right)+\left(1-P_m\right)P_e\right]^i \nonumber\\
\times \left[\left(1-P_m\right)\left(1-P_e\right)+P_m P_e\right]^{K-i}
\end{eqnarray}

The results above were obtained from a direct generalization from \cite{Letaief09}, where a similar expression is derived for the $n=1$ case, and from \cite{letaief09_2} where the overall false alarm and miss-detection probabilities were obtained for the perfect reporting channel case.

Analyzing (\ref{eq:Qf}) and (\ref{eq:Qm}), one can note that if the individual false alarm probability, $P_f$, is not significant, the overall false alarm is given by:
\begin{equation}
\label{eq:Q_infty}
Q_f^\infty\left(n\right)=\lim_{P_f \rightarrow 0}Q_f=\sum_{i=0}^{K-n}{{K}\choose{i}}\left(1-P_e\right)^iP_e^{K-i}
\end{equation}

In a similar way, if the individual miss-detection probability, $P_m$, approaches zero, the overall miss-detection probability is given by:
\begin{equation}
Q_m^{\infty}\left(n\right)=\lim_{P_m \rightarrow 0}Q_m=\sum_{i=K-n+1}^{K}{{K}\choose{i}}P_e^i\left(1-P_e\right)^{K-i}
\end{equation}

We will refer to these probabilities as asymptotic false alarm and miss-detection probabilities, which do not depend on the average SNR $\gamma$ received in the cognitive radio and are completely due to the errors introduced by the report channel.

These asymptotic probabilities, however, depend on the parameter $n$. $Q_f^\infty$ is a decreasing function of $n$, on the other hand, $Q_m^{\infty}$ is a increasing function of $n$. In the next section, we will analyze the system performance dependence on the parameter $n$ choice in some specific scenarios.

\section{Results}
\label{Sec:results}

\begin{figure}
\includegraphics[width=85mm,height=82mm]{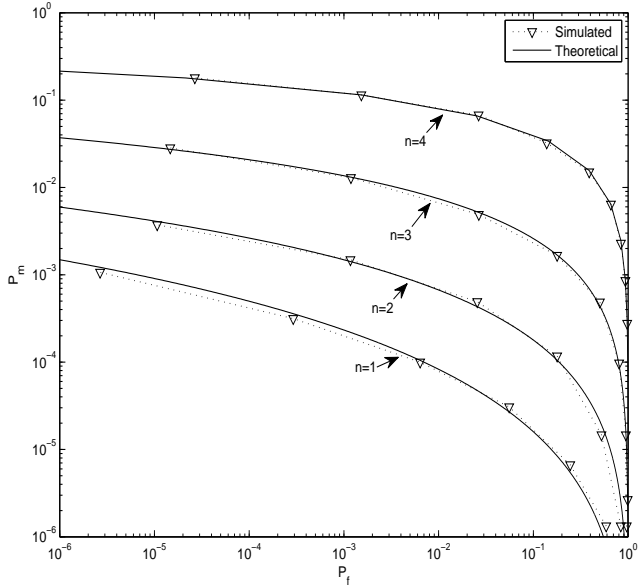}
\caption{ROC - $\gamma=20$dB, $K=4$, perfect reporting channel}
\label{Fig.:ROC_perfect}
\end{figure}

In this section we will describe how to choose the parameter $n$ of a cognitive radio system applying a $n$-out-of-$K$ rule in the fusion center. When the reporting channel is perfect, the $1$-out-of-$K$ rule, \emph{i.e.}, the OR rule, often provides better results \cite{Letaief09}, \cite{ghasemi07}. This fact is attested in the receiver operating characteristics (ROC) curves shown in Fig. \ref{Fig.:ROC_perfect}. In this example, we considerer a cognitive radio system with $K=4$ secondary users, $M=6$ samples and an average SNR $\gamma=20$dB with perfect reporting channel. From Fig. \ref{Fig.:ROC_perfect}, we can observe the system performance for different values of $n$ and conclude that for this situation the performance of the system degrades with increasing $n$.

In the following, we will analyze how the errors introduced by the reporting channel influences the choice of the parameter $n$. We evaluate the same system with ROC depicted in \ref{Fig.:ROC_perfect}, but the SNR in the reporting channel is now given by $SNR_r=10\log_{10}\left(\frac{1}{\sigma^2}\right)=10$dB.

\begin{figure}
\centering
\includegraphics[width=85mm,height=82mm]{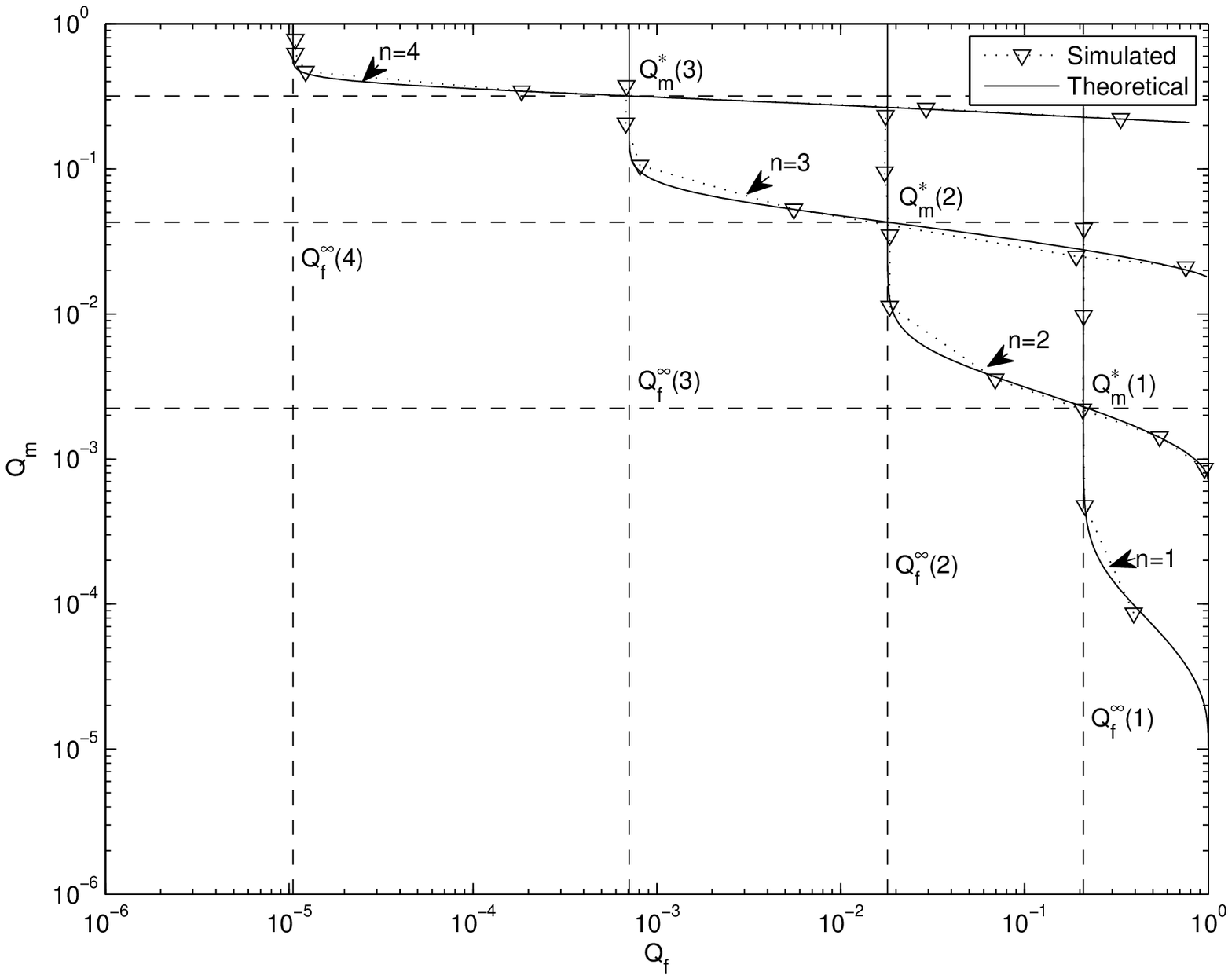}
\caption{ROC - $\gamma=20$dB, $K=4$, $SNR_r=5dB$}
\label{Fig.:ROC_error}
\end{figure}

From Fig. \ref{Fig.:ROC_error}, one can note that decision rule that minimizes the false alarm probability for a given miss-detection probability depends on the miss-detection probability. This is not unexpected since, from eq. (\ref{eq:Q_infty}), one can note that the asymptotic false alarm probability $Q_f^{\infty}$ is a function of $n$. Therefore, for different values of $n$, the minimum achievable false alarm probability is different.

In this situation, the optimum decision rule should be adaptive, depending on the target miss-detection probability. Denoting the target miss-detection probability by $Q_m^t$, we have that the following rule should be applied:

\begin{equation}
n_{opt}=\begin{cases}
1, \quad Q_m^t \leq Q_m^\ast(1)\\
n+1, \quad Q_m^\ast(n) < Q_m^t \leq Q_m^\ast(n+1)\\
K, \quad Q_m^t > Q_m^\ast\left(K-1\right)
\end{cases}
\end{equation} 
where $Q_m^\ast(n)$ corresponds to the minimum miss-detection probability that leads to $Q_f^{\infty}(n)$, as indicated in Fig. \ref{Fig.:ROC_error}.

It is important to emphasize that the optimality criterion is to minimize the false alarm probability for a given target miss-detection probability.

\section{Conclusions}
\label{Sec:conclusions}

It was pointed out throughout this paper that the analysis of the cooperative spectrum sensing system, applying the $n$-out-of-$K$ in the fusion center, should be cautionary when the reporting channel introduces errors. It was shown that, when the reporting errors are take into account, the changes introduced in the ROCs are such that the optimal parameter $n$ is modified.

\bibliographystyle{IEEEbib}
\bibliography{bibli4}

\end{document}